\documentclass[aps,twocolumn,prl,floatfix,superscriptaddress]{revtex4-2}
\usepackage{amsmath,amsfonts,amssymb,color,colordvi,epsfig,graphicx}
\usepackage{bm}
\usepackage{braket}
\usepackage{multirow}
\usepackage{threeparttable}
\usepackage{soul}
\usepackage{xcolor}

\definecolor{dark-gray}{rgb}{.35,.55,.55}
\definecolor{dark-blue}{rgb}{.0,.0,.6}
\usepackage[colorlinks=true,linkcolor=dark-blue,citecolor=dark-blue,urlcolor=dark-blue]{hyperref}

\definecolor{gray}{rgb}{0.75,0.75,0.75}
\definecolor{violet}{rgb}{0.75,0.00,0.75}

\newcommand{\satoyaend}{\color{black}}

\newcommand{\trace}{\operatorname{Tr}}
\newcommand{\one}{\mathbb{I}}

\begin{document}
\title{Experimental verification of bound and multiparticle entanglement with the randomized measurement toolbox
}

\author{Chao Zhang}
\email{These authors contributed equally to this paper.}
\affiliation{CAS Key Laboratory of Quantum Information, University of Science and Technology of China, Hefei 230026, China}
\affiliation{CAS Center for Excellence in Quantum Information and Quantum Physics, University of Science and Technology of China, Hefei 230026, China}
\affiliation{Hefei National Laboratory, University of Science and Technology of China, Hefei 230088, China}

\author{Yuan-Yuan Zhao}
\email{These authors contributed equally to this paper.}
\affiliation{Peng Cheng Laboratory, Shenzhen 518055, China}

\author{Nikolai Wyderka}
\affiliation{Institut für Theoretische Physik III, Heinrich-Heine-Universität Düsseldorf, Universitätsstr.~1, 40225 Düsseldorf, Germany}
\author{Satoya Imai}
\affiliation{Naturwissenschaftlich-Technische Fakultät, Universität Siegen, Walter-Flex-Str.~3, 57068 Siegen, Germany}
\author{Andreas Ketterer}
\affiliation{Fraunhofer Institute for Applied Solid State Physics IAF, Tullastr.~72, 79108 Freiburg, Germany}

\author{Ning-Ning Wang}
\affiliation{CAS Key Laboratory of Quantum Information, University of Science and Technology of China, Hefei 230026, China}
\affiliation{CAS Center for Excellence in Quantum Information and Quantum Physics, University of Science and Technology of China, Hefei 230026, China}
\affiliation{Hefei National Laboratory, University of Science and Technology of China, Hefei 230088, China}

\author{Kai Xu}
\affiliation{CAS Key Laboratory of Quantum Information, University of Science and Technology of China, Hefei 230026, China}
\affiliation{CAS Center for Excellence in Quantum Information and Quantum Physics, University of Science and Technology of China, Hefei 230026, China}
\affiliation{Hefei National Laboratory, University of Science and Technology of China, Hefei 230088, China}

\author{Keren Li}
\affiliation{Peng Cheng Laboratory, Shenzhen 518055, China}

\author{Bi-Heng Liu}
\affiliation{CAS Key Laboratory of Quantum Information, University of Science and Technology of China, Hefei 230026, China}
\affiliation{CAS Center for Excellence in Quantum Information and Quantum Physics, University of Science and Technology of China, Hefei 230026, China}
\affiliation{Hefei National Laboratory, University of Science and Technology of China, Hefei 230088, China}

\author{Yun-Feng Huang}
\email{hyf@ustc.edu.cn}
\affiliation{CAS Key Laboratory of Quantum Information, University of Science and Technology of China, Hefei 230026, China}
\affiliation{CAS Center for Excellence in Quantum Information and Quantum Physics, University of Science and Technology of China, Hefei 230026, China}
\affiliation{Hefei National Laboratory, University of Science and Technology of China, Hefei 230088, China}

\author{Chuan-Feng Li}
\email{cfli@ustc.edu.cn}
\affiliation{CAS Key Laboratory of Quantum Information, University of Science and Technology of China, Hefei 230026, China}
\affiliation{CAS Center for Excellence in Quantum Information and Quantum Physics, University of Science and Technology of China, Hefei 230026, China}
\affiliation{Hefei National Laboratory, University of Science and Technology of China, Hefei 230088, China}

\author{Guang-Can Guo}
\affiliation{CAS Key Laboratory of Quantum Information, University of Science and Technology of China, Hefei 230026, China}
\affiliation{CAS Center for Excellence in Quantum Information and Quantum Physics, University of Science and Technology of China, Hefei 230026, China}
\affiliation{Hefei National Laboratory, University of Science and Technology of China, Hefei 230088, China}

\author{Otfried  G\"uhne}
\email{otfried.guehne@uni-siegen.de}
\affiliation{Naturwissenschaftlich-Technische Fakultät, Universität Siegen, Walter-Flex-Str.~3, 57068 Siegen, Germany}

\date{\today}

\begin{abstract}

In recent years, analysis methods for quantum states based on randomized measurements have been investigated extensively. Still, in the experimental implementations these methods were typically used for characterizing strongly entangled states and not to analyze the different families of multiparticle or weakly entangled states. In this work, we experimentally prepare various entangled states with  path-polarization hyper-entangled photon pairs, and study their entanglement properties using the full toolbox of randomized measurements. First, we successfully characterize the correlations of a series of GHZ-W mixed states using the second moments of the random outcomes, and demonstrate the advantages of this method by comparing it with the well-known three-tangle and squared concurrence. Second, we generate bound entangled chessboard states of two three-dimensional systems and verify their weak entanglement with a criterion derived from moments of randomized measurements.

\end{abstract}

\maketitle

\section{Introduction}
Quantum entanglement is one of the most prominent non-classical features of 
quantum mechanics and often viewed as a resource in quantum information processing \cite{horodecki2009quantum}. Its generation and characterization is of growing interest from both practical and fundamental perspectives. While deciding whether a given quantum state is entangled or not is in general a hard task \cite{gurvits2003classical}, many experimentally feasible schemes exist that verify entanglement in some states.

A prominent example for such schemes are entanglement witnesses, which allow for rather simple detection of entanglement using few measurements, whereas other schemes detect non-locality by evaluating some Bell-type inequalities \cite{guhne2009entanglement}. On the experimental side, numerous entangled states have been generated and multi-qubit entanglement \cite{qubitexp1,qubitexp3}, high-dimensional 
entanglement of two particles \cite{quditexp1,quditexp2,quditexp3}, and also bound entanglement~\cite{boundexp1,boundexp2, boundexp3,boundexp4,boundexp5} has been characterized.

When applying the standard criteria in a practical experiment, however, one always needs to align the local measurement settings strictly or to make some assumptions 
on the target state to prepare, e.g., by tailoring a witness specifically for 
states close to some fixed target state. To remedy this, several schemes based 
on the moments of randomized correlations have been proposed 
\cite{tran2015quantum, tran2016correlations, ketterer2019characterizing, 
ketterer2020entanglement, imai2021bound, ketterer2022statistically, ohnemus2023quantifying, wyderka2023probing, wyderka2022complete, brydges2019, elben2019, elben2020, elben2023, review2023}. They provide an efficient way to characterize multi-particle correlations in states without prior knowledge about the state, nor any alignment of measurement directions. Recently, it has been shown that this approach also allows for the detection of bound entanglement~\cite{imai2021bound}.

In this paper, we implement in a photonic setup the randomized measurement scheme to detect entanglement in mixtures of three-qubit GHZ and W-states using second moments of the random outcomes. Furthermore, we prepare bound entangled chessboard states of two qutrits and show their entanglement by evaluating an
entanglement criterion which is based on the second and fourth
moment of a randomized measurement outcome, without implementing the
random unitaries explicitly. This demonstrates that the criterion
from Ref.~\cite{imai2021bound} is indeed strong enough to capture this weak form of entanglement, even in the presence of noise and experimental imperfections. Our implementation combines the photon's polarization and path degrees of freedom to generate precisely controlled high-dimensional states and demonstrates the versatility and efficiency of the randomized measurement approach.

\section{Theory}\label{sec:theory}

In the randomized measurement scheme \cite{tran2015quantum,tran2016correlations,ketterer2019characterizing, ketterer2020entanglement,imai2021bound,ketterer2022statistically,ohnemus2023quantifying, wyderka2023probing, wyderka2022complete, brydges2019, elben2019, elben2020, elben2023, review2023}, a subset $S\subset\{1,\ldots,n\}$ of the parties of an $n$-partite quantum state $\rho$ of fixed local dimension $d$ is measuring some fixed, local observables in random directions. The moments of the distribution of measurement results can be written as
\begin{align}
    \mathcal{R}_S^{(t)} = \int \text{d}U_1 \ldots \text{d}U_n \langle U_1 \tau_1 U_1^\dagger\otimes\ldots \otimes U_n \tau_n U_n^\dagger\rangle_\rho^t,
\end{align}
where the $\tau_i$ denote the local observables, and $\tau_i = \one$ whenever $i\notin S$.
The integrals are evaluated over the Haar measure of the unitary group $\mathcal{U}(d)$.
In case of qubit systems, one usually chooses $\tau_i = \sigma_z$ for $i\in S$, in which case the second moments ($t=2$) are related to the purities of the reduced states of $\rho$. The sum of second moments for all subsets $S$ of size $\vert S\vert = k$ is proportional to what is known as the $k$-sector length of the state \cite{aschauer2003local,de2011multipartite,klockl2015characterizing,wyderka2020characterizing,eltschka2020maximum,miller2022sector}. In particular, for three qubits the sector lengths $A_k$ are given by
\begin{align}
    A_1&=3(\mathcal{R}_A^{(2)}+\mathcal{R}_B^{(2)}+\mathcal{R}_C^{(2)}),\\
    A_2&=9(\mathcal{R}_{AB}^{(2)}+\mathcal{R}_{AC}^{(2)}+\mathcal{R}_{BC}^{(2)}),\\
    A_3&=27\mathcal{R}_{ABC}^{(2)}.
\end{align}
Decomposing $\rho$ in terms of the local Pauli basis $\{\sigma_0 = \one, \sigma_1 = \sigma_x, \sigma_2 = \sigma_y, \sigma_3 = \sigma_z\}$, yields
\begin{align}
    \rho_{ABC}=\frac{1}{8}\sum_{i,j,k=0}^3 \alpha_{ijk}\sigma_i\otimes\sigma_j\otimes\sigma_k
\end{align}
and allows to express the sector lengths in terms of the coefficients $\alpha_{ijk}$ as follows:
$A_1 = \sum_{i=1}^3 (\alpha_{i00}^2 + \text{perm.})$,
$A_2 = \sum_{i,j=1}^3 (\alpha_{ij0}^2 + \text{perm.})$, and
$A_3 = \sum_{i,j,k=1}^3 \alpha_{ijk}^2$.

 In terms of the sector lengths, several entanglement criteria exist that detect certain entangled states. To proceed, let us recall that a three-particle state $\rho_{ABC}$ is called biseparable for a partition $A|BC$ if
 $ 
 \rho_{A|BC}
     = \sum_k q_k^A \rho_k^A \otimes \rho_k^{BC},
$
where the  positive coefficients $q_k^A$ form a probability distribution. 
Similarly, the biseparable states $\rho_{B|CA}$ and $\rho_{C|AB}$ can be defined.
Moreover, we can consider the mixture of biseparable states for all partitions as
\begin{align}
    \rho_{\text{bisep}}
    = p_A \rho_{A|BC} + p_B \rho_{B|CA} + p_C \rho_{C|AB},
\end{align}
where $p_A, p_B, p_C$ are probabilities.
A quantum state is called genuinely multipartite entangled (GME) if it cannot be written in the form of $\rho_{\text{bisep}}$.
\satoyaend 

For three-qubit states, if $A_3>3$, the state must be GME (the maximal value being $A_3=4$ for the GHZ state $\ket{\text{GHZ}}=\frac{1}{\sqrt{2}}(\ket{000}+\ket{111}$).
A stronger version exists, which states that if
\begin{align} \label{eq:strongbisep}
    A_2 + A_3 > 3(1+A_1),
\end{align}
the state cannot be biseparable w.r.t.~any fixed partition, and strong numerical evidence exists that in that case, even GME states must be present \cite{imai2021bound}.
In this paper, we aim to detect entanglement in a mixture of a GHZ and a W state, given by
\begin{align}
    \rho(g) = g|\text{GHZ}\rangle \langle \text{GHZ}|+(1-g)|\text{W}\rangle \langle \text{W}|,
\end{align}
where $g\in[0,1]$ denotes the amount of mixing and $|\text{W}\rangle = \frac{1}{\sqrt{3}}(\ket{001} + \ket{010} + \ket{100})$.

The family of states $\rho(g)$ exhibits some interesting properties. First, it is supported in the symmetric subspace. This implies that
$F_{XY}\rho(g) = \rho(g)F_{XY} = \rho(g)$,
where
$F_{XY} = \sum_{i,j} \ket{ij}\!\bra{ji}_{XY}$
is the flip (swap) operator acting on the subsystems $XY\in\{AB, BC, CA\}$.
It is known that if a state lives in
the symmetric subspace, it is either fully separable or GME
\cite{eckert2002quantum,ichikawa2008exchange,toth2009entanglement,wei2010exchange,hansenne2022symmetries}.
However, the experimentally generated version of the state $\rho(g)$ cannot be assumed to have the symmetry due to experimental imperfections. Accordingly, the generated state can become biseparable, thus, we employ the criterion in Eq.~(\ref{eq:strongbisep}) to detect its entanglement.
We stress again that the criterion in Eq.~(\ref{eq:strongbisep}) has been conjectured to imply the presence of GME from numerical evidence, but its analytical proof has not yet been provided \cite{imai2021bound}.
That is, even if the criterion Eq.~(\ref{eq:strongbisep}) is verified experimentally, the state may be entangled for any fixed partition, but it can be a mixture of at least three biseparable states for different bipartitions.

Second, when the parameter $g$ is outside the region of $0.297 \leq g \leq 0.612$, the criterion in Eq.~(\ref{eq:strongbisep}) is satisfied.
This parameter region is very close to other well-known regions using two other entanglement measures \cite{lohmayer2006entangled, szalay2011separability}.
On the one hand, the three-tangle $\tau$ vanishes for $0\leq g\leq g_\tau \approx 0.627$, where $\tau$ measures residual (three-partite) entanglement that cannot be expressed as two-body entanglement \cite{coffman2000distributed}.
Note that the GHZ state maximizes the three-tangle, while it vanishes for the W state.
On the other hand, the sum of squared concurrences $C_{A|B}^2 + C_{A|C}^2$ vanishes for $g_C \approx 0.292 \ldots \leq g\leq 1$,
where the concurrence $C_{X|Y}$ measures bipartite entanglement in the reduced state between the parties $X$ and $Y$ \cite{wootters1998entanglement}.
Hence, we can conclude that the criterion in Eq.~(\ref{eq:strongbisep}) can detect the multi-partite entanglement of $\rho(g)$ even in regions where the three-tangle and the concurrence vanish, if the parameter $g$ satisfies
$g_C \leq g < 0.297$ or
$0.612 < g \leq g_\tau$.

In contrast to qubit systems, the second moments of higher-dimensional states are not automatically related to sector lengths. In fact, the choice of the local observables influences which local unitary invariants can be extracted from the moments \cite{wyderka2022complete}. Let us expand a bipartite quantum state of dimension $d$ in terms of some local, hermitian operator basis $\{\lambda_i\}_{i=0}^{d^2-1}$ with $\lambda_0 = \one$, $\trace(\lambda_i\lambda_j) = d\delta_{ij}$, such as the Gell-Mann basis \cite{kimura2003bloch, bertlmann2008bloch, siewert2022orthogonal}. Then
\begin{align}
    \rho = \frac1{d^2}\!\!\Big[\one \otimes \one + \!\! \sum_{i=1}^{d^2-1} \!\! (\alpha_i \lambda_i \otimes \one + \beta_i \one \otimes \lambda_i) + \!\! \sum_{i,j=1}^{d^2-1}  \!\!T_{ij} \lambda_i \otimes \lambda_j\Big]
\end{align}
is called the generalized Bloch decomposition of $\rho$, where the matrix $T$ is known as the correlation matrix of $\rho$. For this matrix, many entanglement criteria exist, most notably the de~Vicente criterion \cite{de2008further}, stating that for separable states, $\trace(\vert T\vert) \leq d-1$. While the left-hand side is not directly accessible from the moments of randomized measurements, it is possible to obtain related quantities by carefully choosing the observables $\tau_i$ as detailed in Ref.~\cite{wyderka2023probing}, such that
\begin{equation}
\begin{aligned}
&\mathcal{R}^{(2)}_{AB}=\text{tr}(TT^\dag)/(d-1)^2\\
&\mathcal{R}^{(4)}_{AB}=\left[\frac{1}{3}\text{tr}(TT^\dag)/(d-1)^2+\frac{2}{3}\text{tr}(TT^\dag TT^\dag)\right]/(d-1)^4.
\end{aligned}
\end{equation}

For example, for $d=3$, $\tau_i = \operatorname{diag}(\sqrt{3/2}, 0, -\sqrt{3/2})$.
The combined knowledge of these two quantities allows to detect entanglement, whenever it is incompatible with the de~Vicente criterion, i.e., if the measured value of $\mathcal{R}^{(4)}_{AB}$ is below the minimum given by
\begin{equation}
\label{eq:3dcriterion}
\begin{aligned}
&\min~\mathcal{R}^{(4)}_{AB}\\
&\text{s.t.}~\mathcal{R}^{(2)}_{AB} = \text{measured}, \text{tr}(|T|)\leq d-1.
\end{aligned}
\end{equation}
Note that this lower bound can also be calculated analytically \cite{wyderka2023probing}.
Interestingly, there exist states which have a positive partial transpose, but can be detected to be entangled by these two moments, implying bound entanglement. A $3\times 3$-dimensional state from the chessboard family of bound entangled states described in Ref.~\cite{bruss2000construction} (also see Appendix C2 in \cite{imai2021bound}) has been identified to violate it extremely, which makes it a good candidate to prepare and detect its entanglement experimentally. It is given by
\begin{equation}\label{chessboardS}
\rho_{\text{ch}}=N\sum_{i=1}^4\ket{V_i}\bra{V_i}, 
\end{equation}
where $N=1/\sum_i\langle{V_i}|{V_i}\rangle^2=1/4$ 
is a normalization factor and
\begin{equation}
    \begin{split}
        &\ket{V_1}=1/\sqrt{6}(\ket{0}+2\ket{2})\ket{0}+1/\sqrt{6}\ket{11},\\
        &\ket{V_2}=1/\sqrt{6}(-\ket{0}+2\ket{2})\ket{1}+1/\sqrt{6}\ket{10},\\
        &\ket{V_3}=1/\sqrt{6}\ket{0}(-\ket{0}+2\ket{2})+1/\sqrt{6}\ket{11},\\
        &\ket{V_4}=1/\sqrt{6}\ket{1}(\ket{0}+2\ket{2})+1/\sqrt{6}\ket{01}.
    \end{split}
\end{equation}

\section{Experimental setup}\label{sec:expsetup}

\begin{figure}[t]
\centering
\includegraphics[width=0.99\linewidth]{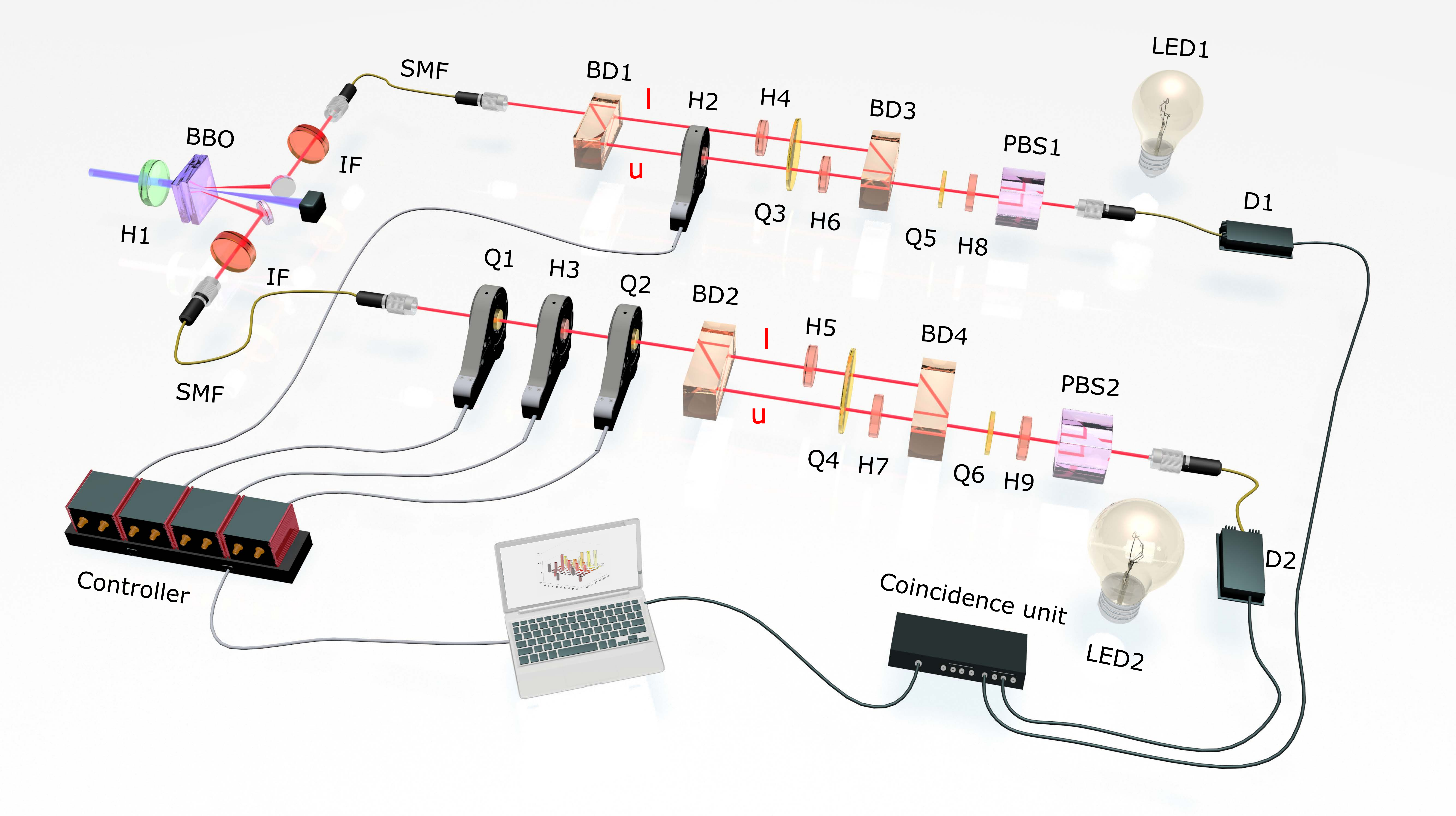}
    \caption{Experimental setup for the chessboard state. 
    The hyper-entangled state $\ket{\psi_s}=\sqrt{5/6}\ket{00}+\sqrt{1/6}\ket{11}$ is prepared first, and the pseudo-random numbers generated from a computer control the angles of the motorized wave plates Q1, Q2, H2, and H3 in order to transform the state to $\ket{V_i}$ randomly. In  the end, the photon pairs are detected by the detectors D1 and D2, and the coincidences are recorded by the coincidence unit ID 800. See text for more details.}
    \label{fig:setup1}
 \end{figure}

We proceed with a description of the experimental implementation. The GHZ-W mixed states are prepared by resorting to the states entangled in polarization degree of freedom (d.o.f.) and path d.o.f. of the photon (that is, hyper-entangled) and with methods similar to the ones in Refs.~\cite{Osetup2015,Osetup2021}. More detailed information about the state preparation of this family of states is given in Appendix~A. 

When preparing the bound entangled chessboard state, it is important to ensure that all its eigenvalues remain non-negative under partial transposition. However, the chessboard state is not of full rank. Affected by the imperfections of 
the experiment, slightly negative eigenvalues of the partial transposition are likely to appear. A more robust way is to prepare 
the state with a level of white noise \cite{boundexp2},
\begin{equation}
\rho_\text{ch}(p)=(1-p)\rho_\text{ch}+p\dfrac{\mathbb{I}}{16}. 
\end{equation}

First, let us briefly review the state preparation procedure.  
As depicted in Fig.~\ref{fig:setup1}, we generate polarization 
entangled ($2\times2$ entangled) photon pairs through a spontaneous 
parametric down-conversion (SPDC) process. Subsequently, we expand the dimensionality of the system by introducing the path modes $u$ and $l$. This will results in three modes: $H_u$, $V_u$, and $H_l$, where ${H}_u$ represents a horizontally polarized photon occupying path $u$, and so on. Finally, specific operations are applied to the system to steer the state to the target ones.

Specifically,  a Half-Wave Plate (HWP) H1 with the optic axis placed at $12.05^\circ$ is used to rotate a 390 nm horizontally polarized pump laser (with an 80 MHz repetition rate and a 140-fs pulse duration)  to state $\ket{\psi_p}=\sqrt{5/6}\ket{H}+\sqrt{1/6}\ket{V}$, where $H$ and $V$ represent the horizontal and the vertical polarization, respectively. The pump photon is then split into two photons after pumping two crossed-axis type-I $\beta$-Barium Borate (BBO) crystals in the SPDC process, transforming the state into $\ket{\psi_p}\rightarrow\sqrt{5/6}\ket{HH}+\sqrt{1/6}\ket{VV}$. By passing through the Beam Displacers (BDs) BD1 and BD2, the down-converted photons' $H-$($V-$) components are directed to path $u$ ($l$). And for path mode $u$, we have the mode labeled as $H_u$ and $V_u$. By re-encoding $\ket{H}_u\rightarrow\ket{0}$, $\ket{V}_l\rightarrow\ket{1}$, and $\ket{V}_u\rightarrow\ket{2}$, we obtain the hyper-entangled state $\ket{\psi_s}=\sqrt{5/6}\ket{H_uH_u}+\sqrt{1/6}\ket{V_lV_l}\rightarrow\sqrt{5/6}\ket{00}+\sqrt{1/6}\ket{11}$. 

It is worth noting that all the four states $\ket{V_i}$ in Eq.~(\ref{chessboardS}) can be generated by performing local 
operations on the state $\ket{\psi_s}$,  
\begin{equation}
    \begin{split}
        &{\ket{V_1}}=U_2\otimes \mathbb{I}\ket{\psi}, 
        \quad
      {\ket{V_2}}=U_3\otimes U_1\ket{\psi},
        \\
        &{\ket{V_3}}=\mathbb{I}\otimes U_3\ket{\psi}, \quad
        {\ket{V_4}}=U_1\otimes U_2\ket{\psi},
    \end{split}
\end{equation}
where
\begin{equation}
\begin{aligned}
&U_1=\left (\begin{matrix}
0 & 1 & 0 \\
1 & 0 & 0 \\
0 & 0 & 1 
\end{matrix}
\right ),\quad
U_2=\left (\begin{matrix}
\sqrt{1/5} & 0 & \sqrt{4/5} \\
0 & 1 & 0 \\
\sqrt{4/5} & 0 & -\sqrt{1/5} 
\end{matrix}
\right ), \\
&U_3=\left (\begin{matrix}
-\sqrt{1/5} & 0 & \sqrt{4/5} \\
0 & 1 & 0 \\
\sqrt{4/5} & 0 & \sqrt{1/5} 
\end{matrix}
\right ).\\
\end{aligned}
\end{equation}
For the states $\ket{V_3}$ and $\ket{V_4}$, it also works by applying the unitary $U_3\otimes\mathbb{I}$, and $U_2\otimes U_1$, respectively, and then exchanging the labels for the two detectors D1 and D2. Therefore, through performing the operator $U_3$ or $U_2$ on one photon of a pair and the operator $U_1$ or $\mathbb{I}$ on the other photon simultaneously, the state $\ket{\psi_s}$ will be transformed to each of the four states $\ket{V_i}$. The switches between these operators are implemented by the motorized rotating HWPs and Quarter-Wave Plates (QWPs), which are controlled by the pseudo-random numbers generated from a classical computer. Two adjustable LED lights are placed before the detectors to introduce the different levels of white noise into the system.

In the measurement part, a QWP and an HWP located at path $u$ are used to analyze the correlations between basis elements $\ket{0}$ and $\ket{2}$, and now the afterward BD works as a Polarization Beam Splitter (PBS). When measuring the superposition of basis elements $\ket{0}$ and $\ket{1}$, as well as $\ket{2}$ and $\ket{1}$, we first convert the path d.o.f.~to the polarization d.o.f.~via the wave plates and BDs, and then analyze with the combination of the QWP and the HWP. Detailed settings of the wave plates for standard quantum state tomography are given in Tab.~\ref{tab:basis} of Appendix~B. For each measurement basis, we randomly change the photon states to every one of the four states $\ket{V_i}$. The two-photon coincidence counts are recorded per 10 s.

When it comes to measuring the randomized correlations, as elaborated in the theoretical framework,  two distinct approaches are considered. The first one involves conducting local randomized measurements, while the second entails the direct application of Pauli operators or Gell-Mann matrices. In this study, we thoroughly examine and contrast these two methodologies for three-qubit states, utilizing a LabVIEW program to facilitate the automation of numerous measurements. Further details regarding the randomized measurement techniques can be found in the  Appendix~C. For the bound entangled states, we opt to directly measure the 81 combinations of Gell-Mann matrices to avoid the systematic errors that may emerge from the construction of $3\times 3$ random unitaries.

\section{Results}\label{sec:results}

\subsection{Results for the GHZ-W mixed states}

\begin{figure}[t]
\centering
\includegraphics[width=0.99\linewidth]{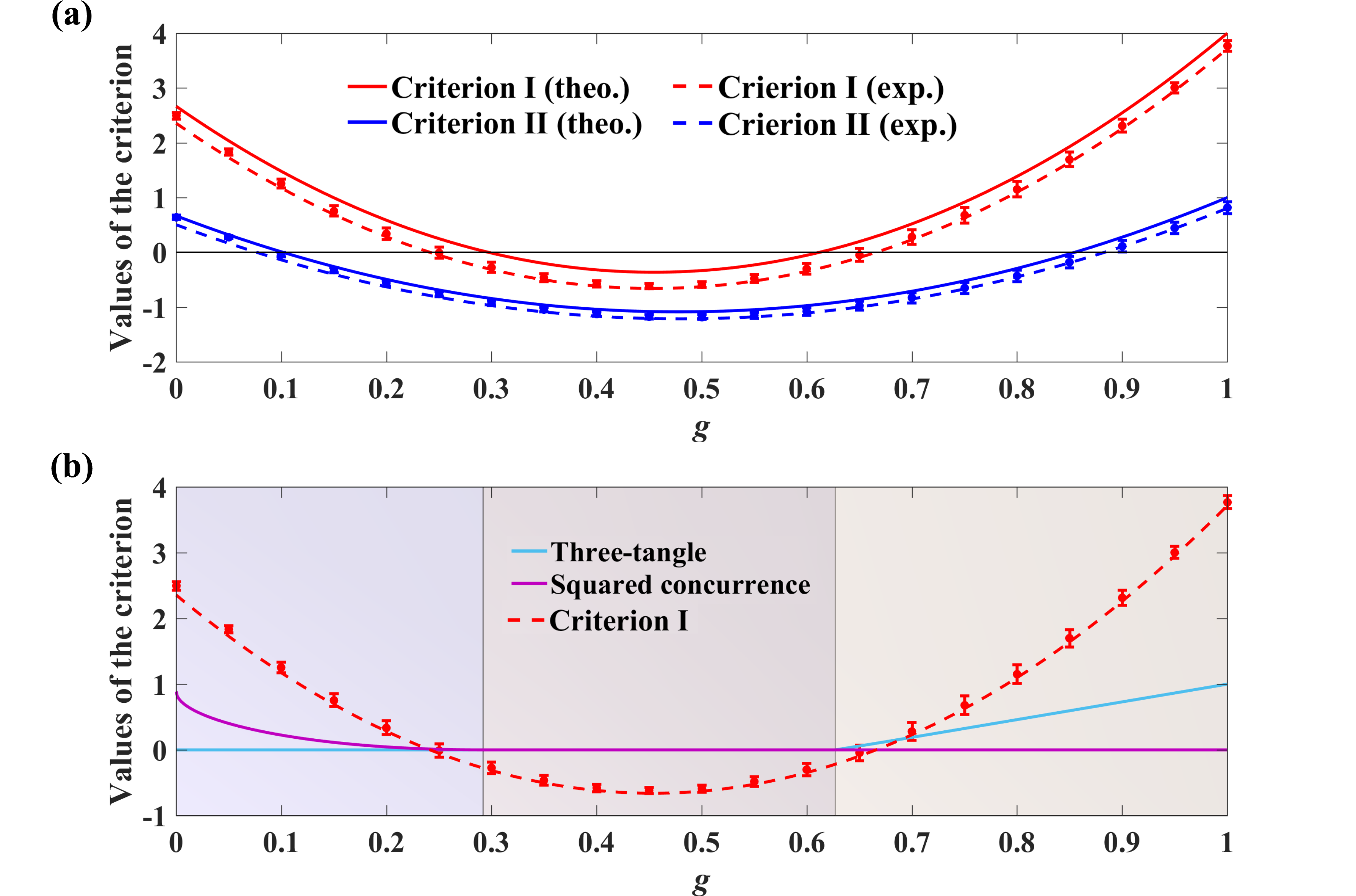}
    \caption{Entanglement analysis via randomized correlations (a) and comparison with other criteria (b). Solid lines: Values of the entanglement Criterion I (red), Criterion II (blue),  3-tangle (cyan), and the squared concurrence (purple) for an ideal $\rho(g)$. Dashed lines: Numerical values of the entanglement Criterion I (red) and Criterion II (blue) calculated from the coefficients $\alpha_{ijk}$ of state $g\rho_\text{GHZ}^\text{exp}+(1-g)\rho_\text{W}^\text{exp}$. Dots: Experimental values of Criterion I (red) and Criterion II (blue) obtained from randomized measurements. Here, Criterion I and  Criterion II represent the entanglement criterion $A_2+A_3-3(1-A_1)>0$  and $A_3-3>0$, respectively. The violet color and the light salmon color denote regions where $\rho(g)$ has no three-tangle and no squared concurrence. }
    \label{fig:result1}
  \end{figure}

In our experiment, a set of  GHZ-W mixed states $\rho(g)$ with step size $0.05$ is prepared. For each state, $4000$ measurements in randomized directions are performed, and for each measurement, about $5300$ copies of the state are detected. 
The entanglement criterion of Eq.~(\ref{eq:strongbisep}) is calculated from the randomized measurement data with the error bars obtained by repeating the whole process ten times. From the results in Fig.~\ref{fig:result1}(a), we see that for $0\leq p\leq 0.2$ and $0.7\leq p\leq 1$, the criterion in Eq.~(\ref{eq:strongbisep}) is violated, while the criterion $A_3-3\leq 0$ is not. Clearly, Eq.~(\ref{eq:strongbisep}) improves the previous one. 

Note that the sector length $A_k$ can also be expressed in terms 
of the coefficients $\alpha_{ijk}$, and then compared with 
the randomized measurements. Resorting to the standard 
quantum state tomography process, we obtain the density matrix of the GHZ state $\rho_\text{GHZ}^\text{exp}$ and W state $\rho_\text{W}^\text{exp}$, respectively.
The values of the criterion of Eq.~(\ref{eq:strongbisep}) are calculated from the state $\rho(g)=g\rho_\text{GHZ}^\text{exp}+(1-g)\rho_\text{W}^\text{exp}$ and plotted as the dashed red lines in Fig.~\ref{fig:result1}(a) and (b). 
In contrast, for the ideal states, we have $(A_1, A_2, A_3)=(\frac{(1-g)^2}{3}, 8g^2-8g+3, 4g^2+\frac{11(1-g)^2}{3})$, and the theoretical values of the criteria are shown as the solid red lines in Fig.~\ref{fig:result1}. 

We see that the results deduced from randomized measurements and from the coefficients $\alpha_{ijk}$ are approximately identical, providing evidence for the correct implementation of the randomized measurements. In the region $0.08\leq g\leq 0.24$ and $0.67\leq g\leq 0.88$, where the criterion $A_3-3\leq 0$ fails, we detect genuinely multi-partite entanglement. Furthermore, from
Fig.~\ref{fig:result1}(b), we see that our criterion still works for $g\leq 0.24$ in the violet color region where the states have no three-tangle and also for $g\geq0.67$ in the light salmon region where they exhibit no squared concurrence.

\subsection{Results for the chessboard state}

\begin{figure}[t]
\centering
\includegraphics[width=0.9\linewidth]{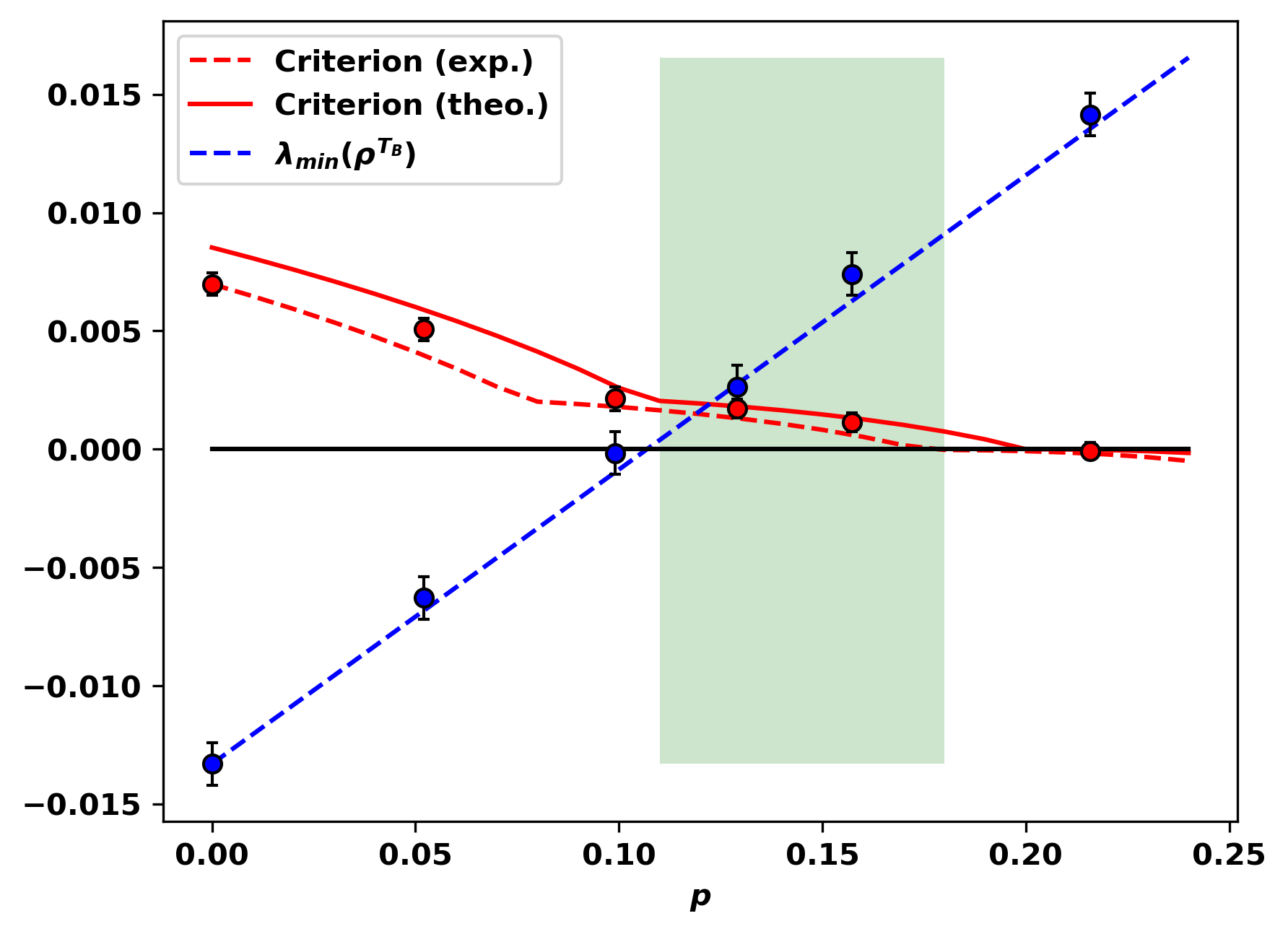}
    \caption{Smallest eigenvalue of the partial transposition and the value of the entanglement criterion in Eq.~(\ref{eq:3dcriterion}), evaluated from the tomographic data of noisy chessboard states for different noise levels $p$. Red, solid: Value of the entanglement criterion for the ideal noisy chessboard state. Red, dashed: Criterion for the experimental noiseless chessboard state mixed with ideal white noise. Red dots: Value of the criterion for experimental noisy states for different noise levels. Blue, dashed: Smallest eigenvalue of the partial transposition of the experimental noiseless chessboard state mixed with ideal white noise. Blue dots: smallest eigenvalue of partial transposition of the experimental noisy states. The green region shows the range where bound entanglement is detected.}
    \label{fig:chessboard}
  \end{figure}

The experimentally prepared chessboard state $\rho_\text{ch}^\text{exp}$ is reconstructed using the maximum-likelihood algorithm.  Due to imperfections, when no white noise is added, the minimal eigenvalues of the partially transposed (PT) density matrix is $-0.0133$, such that state is not PPT and probably not bound entangled. To remove these negative eigenvalues, we introduce different levels of white noise between $p=0$ and $p=0.22$ in the experiment, and plot the minimum PT eigenvalue and the violation of the entanglement criterion in Eq.~(\ref{eq:3dcriterion}) in Fig.~\ref{fig:chessboard}. In particular, for the state with noise level $p=0.1291$, the minimum PT eigenvalue equals $0.0026\pm0.0009$ and the fidelity between the experimentally prepared state $\rho_\text{ch}^\text{exp}$ and the the noisy chessboard state $\rho_\text{ch}(p=0.1291)$ is given by $F(\rho_\text{ch},\rho_\text{ch}^\text{exp})=\text{tr}\big(\sqrt{\sqrt{\rho_\text{ch}}\rho_\text{ch}^\text{exp}\sqrt{\rho_\text{ch}}}\big)=0.9893\pm 0.0012$.

Next, we show that the state is entangled by using the tool of the second and fourth moments. For the state under consideration at $p=0.1291$, the second moment is given by $\mathcal{R}^{(2)}_{AB}=0.2355\pm0.0015$, and the fourth moment by $\mathcal{R}^{(4)}_{AB}=0.0259\pm0.0003$, while for separable states, the lower bound on the fourth moment is given by $0.0277$ for $\mathcal{R}^{(2)}_{AB}=0.2355$ when performing the optimization program in Eq.~(\ref{eq:3dcriterion}). We see that the experimental value $0.0259$ is smaller than the lower bound $0.0277$ and violates it with $6$ standard deviations. Therefore, we experimentally prepared a $3\times3$ bound entangled state with the photonic platform and analyzed its entanglement property via the second and fourth moments 
successfully.

\section{Conclusion}

We experimentally produced a variety of genuinely entangled photonic states consisting of entangled photon pairs amended with path degrees of freedom and characterized them using methods based on locally randomized measurements.  First, we showed how to generate genuinely entangled states of three parties and verified them using entanglement criteria based only on the second moments of the randomized measurements. The latter enabled the verification of mulitpartite entanglement in regimes where well-known measures of multipartite entanglement, i.e., the three-tangle or the squared concurrence, are zero. Further on, we demonstrated the production of weakly bound entangled chessboard states of two qutrits and used entanglement criteria based on the second and fourth moments of the taken randomized measurements to analyze the produced states. As a result, bound entangled states with mixed-state fidelities beyond $98\%$ were successfully produced and verified.

Our work demonstrates the outstanding control of quantum states 
in photonic setups and presents an efficient way for preparing a
low-rank bound entangled state. By incorporating appropriate white 
noise, the setup demonstrates increased robustness against 
transitioning into the free entangled region. Compared with several previous experiments, the precise control allowed us to directly verify 
bipartite bound entanglement in minimal case of a $3\times3$ system, 
without resorting to the various forms of bound entanglement in higher dimensions or in multiparticle systems. This will facilitate further exploration of interesting entanglement effects in
experiments.

\section*{Acknowledgements}
We thank Xiao-Dong Yu for discussions. The work in USTC is supported by the National Natural Science Foundation of China (Nos. 11821404, 11734015, 62075208), the Fundamental Research Funds for the Central Universities (Nos. WK2030000061, YD2030002015), and the Innovation Program for Quantum Science and Technology (No. 2021ZD0301604). Y.Z. is support by the Major Key Project of PCL. S.I. and O.G. are supported by the Deutsche Forschungsgemeinschaft (DFG, German Research Foundation, project numbers 447948357 and 440958198), the Sino-German Center for Research Promotion (Project M-0294), the ERC (Consolidator Grant 683107/TempoQ), and the German Ministry of Education and Research (Project QuKuK, BMBF Grant No.~
16KIS1618K). S.I. acknowledges the support from the DAAD.  N.W.~acknowledges support by the QuantERA project QuICHE via the German Ministry of Education and Research (BMBF
Grant No.~16KIS1119K).

\newpage
\onecolumngrid
\appendix
\section{Appendix A: Experimental details on the preparation of the GHZ-W mixed states}\label{sec:GHZWstatesprepare}

\begin{figure}[h]
\centering
\includegraphics[width=0.9\linewidth]{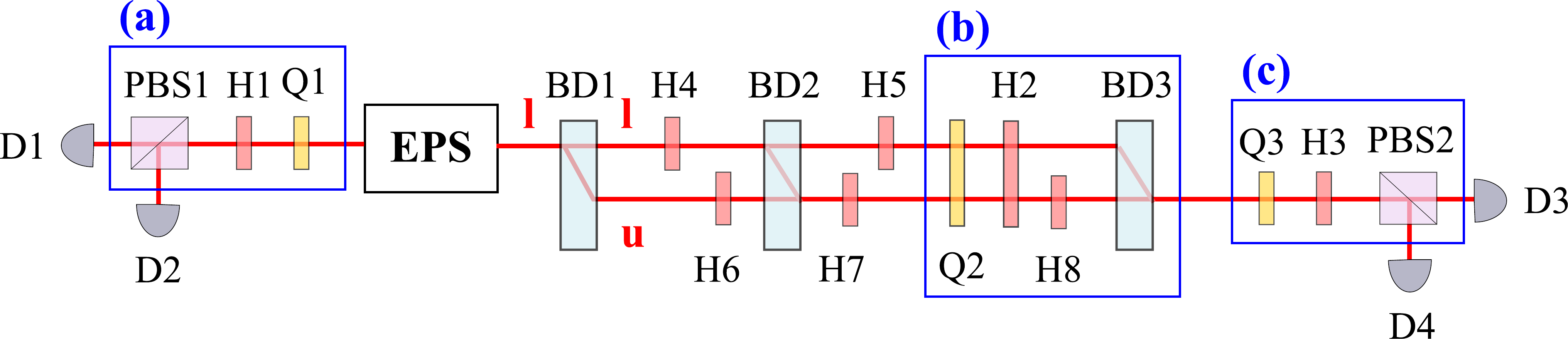}
    \caption{Diagram of the experimental setup for GHZ-W mixed states. See text for further details.}
    \label{fig:setup2}
\end{figure}
In our experiment, the GHZ-W mixed states are prepared using the setup shown in Fig.~\ref{fig:setup2}, and the switch between the GHZ state and W state is realized by engineering the polarization-entangled photon source (EPS), and the subsequent unitary transformations constituted by Beam Displacers (BDs) and the Half-Wave Plates (HWPs). First, for the GHZ state, a polarization-entangled state $\ket{\psi_s}=1/\sqrt{2}(\ket{HH}+\ket{VV})\ket{l}$ is generated through the type-I Spontaneous Parametric Down-Conversion (SPDC) process, and $\ket{u}$ ($\ket{l}$) in Fig.~\ref{fig:setup2} represents the path $u$ (path $l$). Then, BD1 makes the vertically polarized part of the light passes through directly to path $l$, while the horizontal component passes with a 4~mm deviation to path $u$. That is to say, the BD1 performs as a CNOT gate with the polarizations as the controlled qubit and the path as the target qubit. When we set the angles of the half-wave plates H4$\sim$H5 as $0^\circ$ and H6$\sim$H7 as $45^\circ$, we get $\ket{\psi_s}\rightarrow1/\sqrt{2}(\ket{HH}\ket{u}+\ket{VV}\ket{l})$. By encoding the $H$ ($u$) and $V$ ($l$) to the logic qubit $0$ and $1$, we prepare the system into the three qubit GHZ state $\ket{\text{GHZ}}=1/\sqrt{2}(\ket{000}+\ket{111})$.

When it comes to the W state, the EPS is tuned to the state $\ket{\psi_s}=1/\sqrt{3}\ket{VH}\ket{l}+\sqrt{2/3} \ket{HV}\ket{l}$ by rotating the polarization directions of the pump beam to $\ket{\psi_p}=1/\sqrt{3}\ket{H}+\sqrt{2/3}\ket{V}$ and performs a bit flip operation on one of each paired photon generated in the SPDC process. Now the angle of H4 is placed at $-67.4^\circ$ and the one of H5 at $45^\circ$ to transform the state $\ket{V}\ket{l}$ to $1/\sqrt{2}(\ket{V}\ket{u}+\ket{H}\ket{l})$, and $\ket{\psi_s}\rightarrow1/\sqrt{3}\ket{VH}\ket{u}+1/\sqrt{3} \ket{H}\ket{V}\ket{u}+1/\sqrt{3}\ket{H}\ket{H}\ket{l}$. With re-encoding,  the W state $\ket{\text{W}}=1/\sqrt{3}(\ket{100}+\ket{010}+\ket{001})$ is generated. 

At last, various states $\rho(g)=g|\text{GHZ}\rangle \langle \text{GHZ}|+(1-g)|\text{W}\rangle \langle \text{W}|$ are generated by randomly switching the settings of the setup to produce state $\ket{\text{GHZ}}$ or $|\text{W}\rangle$, with probabilities $g$ and $1-g$, respectively.

In the measurement stage, the combination of a Quarter-Wave Plate (QWP), an HWP, and a Polarization Beam Splitter (PBS) enables the polarization state measurement in an arbitrary basis. Thus, the two polarization encoded qubits are analyzed with the devices boxed as parts (a) and (b), respectively. Here BD3 combined with H8 performs as a PBS with only one output port, so we must rotate Q2 and H2 twice to realize the projective measurements $\{U\ket{0}\bra{0}U^\dag, U\ket{1}\bra{1}U^\dag\}$. The third qubit, i.e., the path qubit, is transformed to the polarization degree of freedom, and then analyzed by wave plates Q3, H3, and PBS2 in the boxed part (c).

To facilitate the massive randomized measurements, i.e., 40,000 sets for each state $\rho(g)$ in our experiment, the QWPs Q1$\sim$Q3 and HWPs H1$\sim$H3 are all mounted in Motorized Rotation Mounts (Newport, CONEX-PR50CC). For each local measurement setting drawn uniformly at random, a classical computer inputs the corresponding settings of the QWP and HWP and controls the wave plates automatically rotated to the target angles to perform the measurement.  This entire process is executed via a LabVIEW program.

\begin{figure}
\centering
\includegraphics[width=0.9\linewidth]{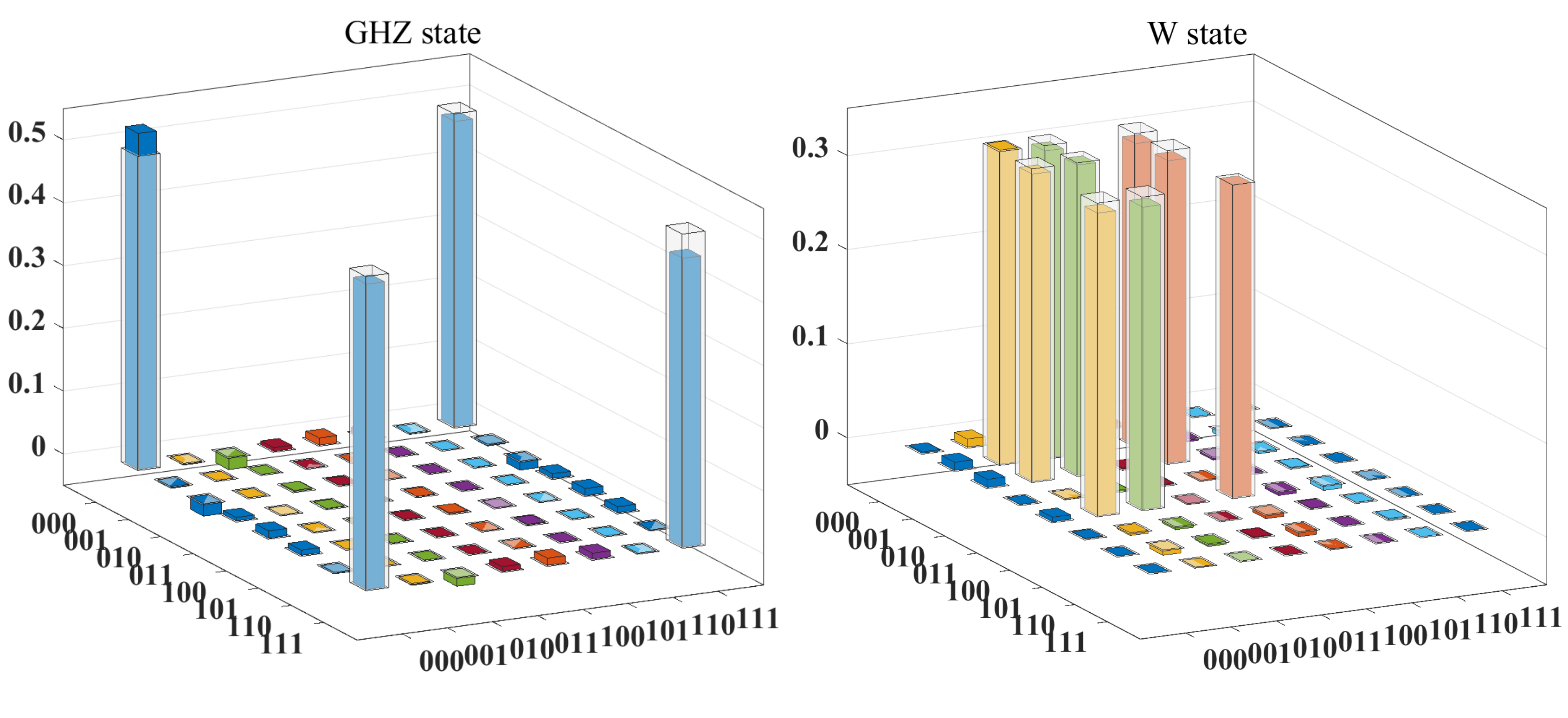}
    \caption{Quantum state tomography for GHZ state and W state. The real parts of the experimentally prepared state are shown as the colored bars, and the corresponding theoretical values are displayed as transparent bars. Each state is constructed from about 2\,700\,000 photon pairs. }
    \label{tomo3qubit}
\end{figure}

Here the quality of the state $\rho(g)$ depends heavily on the GHZ state and the W state, so we give the benchmarks of these two states through quantum state tomography. We estimate the fidelities of the experimentally prepared state and the ideal state $F(\rho^\text{ideal},\rho^\text{exp})=\left(\text{tr}\sqrt{\sqrt{\rho^\text{ideal}}\rho^\text{exp}\sqrt{\rho^\text{ideal}}}\right)$ are $0.9919$ and $0.9890$ for GHZ state and W state, respectively. The real parts of the experimentally prepared state are shown in Fig.~\ref{tomo3qubit}. All fidelities of the GHZ-W mixed states shown as the dots in Fig.~\ref{fig:result1} are above $0.9836$, which shows the good performance of the setup. The error bars are of the size of about $0.0001$, which is obtained with Monte Carlo simulations by sampling the experimentally collected data. 

\section{Appendix B: Quantum state tomography for the chessboard state}\label{sec:tomoCh}

As the red points in  Fig.~\ref{fig:chessboard} show, various noisy chessboard states $\rho_{\text{ch}}(p)$ are prepared to study their entanglement properties. Here, the level of white noise $p$ is estimated by comparing the total coincidence counts with the counts recorded when no white noise source is added, i.e., when the LED lights in Fig.~\ref{fig:setup1} are turned off. For instance, if we record a total of  photonic counts $N_p$  for state $\rho_{\text{ch}}(p)$  and $N_0$ for state with no added white noise, then $p$ is set to the value of $1-N_0/N_p$.

To characterize the chessboard state that we prepared experimentally, we perform a standard quantum state tomography process, where the 81 vectors
$\ket{u_i}\otimes\ket{u_j}$ ($i, j=0,1,...8$) are measured. The detailed forms of the kets $\ket{u_i}$ are given by
\begin{equation}
\begin{split}
&\ket{u_0}=\ket{0};
\ket{u_1}=\ket{1};
\ket{u_2}=\ket{2};\\
&\ket{u_3}=(\ket{0}+\ket{1})/\sqrt{2};
\ket{u_4}=(\ket{0}+i\ket{1})/\sqrt{2};\\
&\ket{u_5}=(\ket{1}+\ket{2})/\sqrt{2};
\ket{u_6}=(\ket{1}+i\ket{2})/\sqrt{2};\\
&\ket{u_7}=(\ket{0}+\ket{2})/\sqrt{2};
\ket{u_8}=(\ket{0}+i\ket{2})/\sqrt{2}.
\end{split}
\end{equation} 
Each basis is realized with the settings in Tab.~\ref{tab:basis}. 

\begin{table}[t]
    \centering
    \begin{threeparttable}
    \begin{tabular}{c||c|c|c|c|c}
    & H4(H5) & Q3(Q4) & H6(H7) & Q5(Q6) &  H8(H9)\\
    \hline
    \hline
    $\ket{u_0}$ & NR & 0 & $45^\circ$ & 0 & $45^\circ$ \\
    \hline
    $\ket{u_1}$ & $45^\circ$ & NR & NR & 0 & 0 \\
    \hline
    $\ket{u_2}$ & NR & 0 & 0 & 0 & $45^\circ$ \\
    \hline
    $\ket{u_3}$ & $45^\circ$ & 0 & $45^\circ$ & $45^\circ$ & $22.5^\circ$ \\
    \hline
    $\ket{u_4}$ & $45^\circ$ & 0 & $45^\circ$  & $90^\circ$ & $22.5^\circ$  \\ 
     \hline
    $\ket{u_5}$ & $45^\circ$ & 0 & 0 & $45^\circ$ & $22.5^\circ$ \\
    \hline
    $\ket{u_6}$ & $45^\circ$ & 0 & 0 & $90^\circ$ & $22.5^\circ$ \\
     \hline
    $\ket{u_7}$ & NR & $45^\circ$ & $22.5^\circ$ & 0 & $45^\circ$ \\
     \hline
    $\ket{u_8}$ & NR & $90^\circ$ & $22.5^\circ$ & 0 & $45^\circ$ \\
    \end{tabular}
    \begin{tablenotes}
    \footnotesize
    \item[*] NR: No Restriction.
    \end{tablenotes}
    \end{threeparttable}
    \caption{The settings of the wave plates to realize the measurements $\ket{u_i}$.}
    \label{tab:basis}
\end{table}

We get the fidelities $0.9835\pm0.0005$, $0.9838\pm0.0006$, $0.9853\pm0.0005$, $0.9893\pm0.0012$, $0.9911\pm0.0005$, $0.9930\pm0.0003$ for states of $p=0,0.052,0.0991,0.1291,0.1573,0.2158$, respectively. The error bars are estimated with  Monte Carlo simulations by sampling the experimental data 100 times. 

\begin{figure*}[t]
\centering
\includegraphics[width=0.99\linewidth]{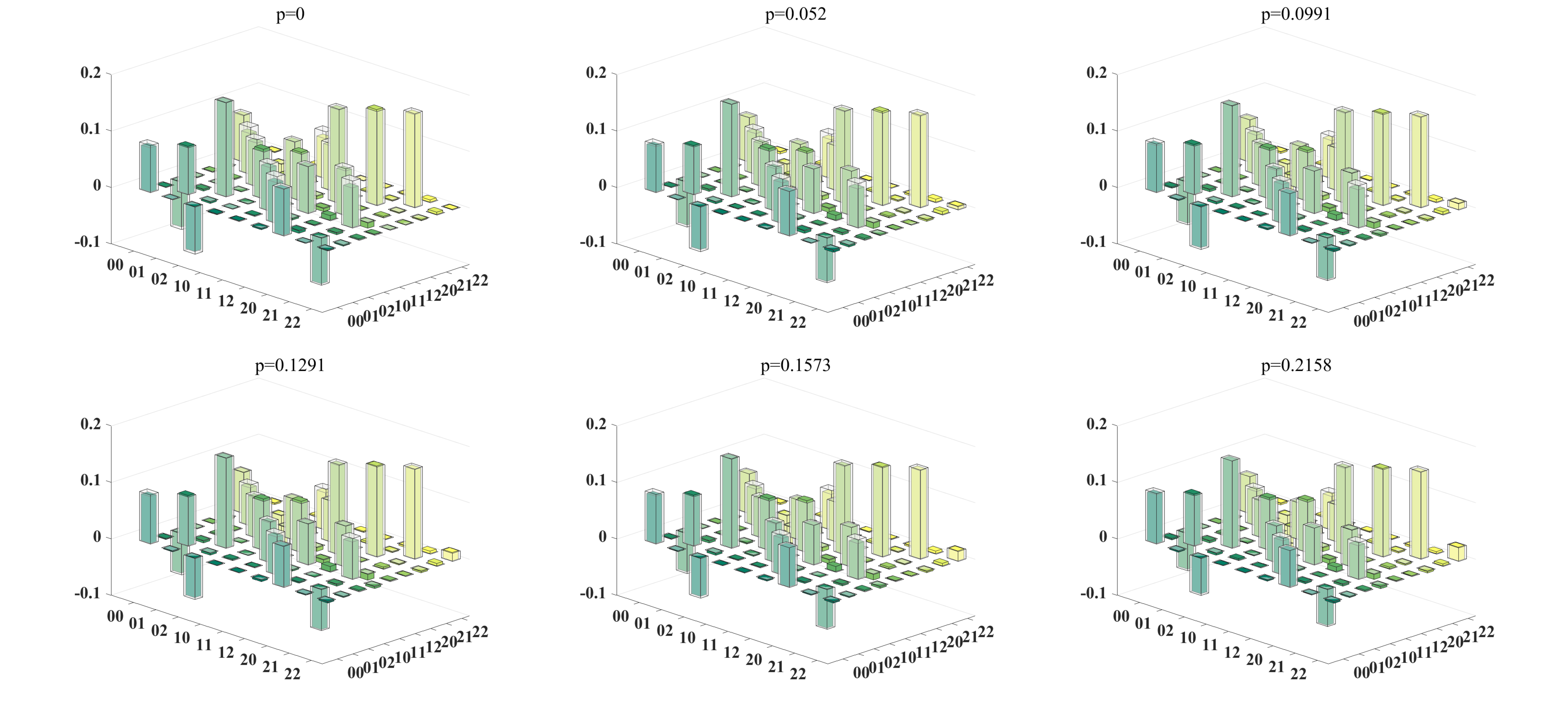}
    \caption{Tomographic reconstruction for chessboard states with varied levels of white noise. The transparent bars are shown as the correspondingly theoretical values of the basis.}
    \label{fig:chessboardstates}
\end{figure*}

\section{Appendix C: Entanglement detection for three-qubit states with randomized measurements}\label{sec:3qubitRM}

In our work, we use the criterion based on the second moment,
\begin{align}
    \mathcal{R}_S^{(2)} = \int \text{d}U_1 \ldots \text{d}U_n \langle U_1 \tau_1 U_1^\dagger\otimes\ldots \otimes U_n \tau_n U_n^\dagger\rangle_\rho^2, 
\end{align}
to study the entanglement property of the three-qubit state $\rho(g)$, where $\tau_i=\sigma_z$ for $i\in S$ and $\tau_i=\one$ for $i\notin S$. 

As each observable $\tau_i$ is measured in the standard basis $\ket{0}$ and $\ket{1}$, we will sort the detection outcomes into eight categories corresponding to the eight basis states $M_{\text{ABC}}=\{\ket{000}\bra{000}, \ket{001}\bra{001}, \ket{010}\bra{010}, \ket{011}\bra{011},\\ \ket{100}\bra{100}, \ket{101}\bra{101}, \ket{110}\bra{110}, \ket{111}\bra{111}\}$, respectively. In every single trial, instead of preparing the state $\rho_U=U\rho(g) U^\dagger$ and then making measurements in the standard basis, we directly perform the measurements $U^\dagger M_{\text{ABC}}U$ on the state $\rho(g)$ in our experiment, where $U=U_A\otimes U_B\otimes U_C$. These two ways are equivalent to each other.

For each choice of local unitaries, we prepare $N$ copies of the state to estimate the probability distributions of the outcomes, and a total of $M$ random unitaries are applied to form the average over local unitaries.

We note that given the observable $\tau_i$ we choose, there are only two possible outcomes $X_i\in \{1, -1\}$ for $\tau_{ABC}=\tau_1\otimes\tau_2\otimes\tau_3$. We define the probability for each outcome as $p_i$, which can be obtained by summing up the probabilities that correspond to the same measurement outcomes. As an example, consider the moment $\mathcal{R}_A^{(2)}$, then $\tau_1=\sigma_z$, $\tau_2=\one$, and $\tau_3=\one$, the outcomes assigned to the eight basis states $M_{ABC}$ are $1,1,1,1,-1,-1,-1,-1$, respectively. We get the probabilities $p_{1}=
p_{000}+p_{001}+p_{010}+p_{011}$ and $p_{2}=p_{100}+p_{101}+p_{110}+p_{111}$, where $\{p_{1}, p_{2}\}$ represents the probability distribution for outcomes $\{1, -1\}$, and $p_{000}=\langle 000|\rho_U|000\rangle$ etc. 

Next, we need to construct the unbiased estimator for $\text{Tr}(\rho U\tau_{ABC}U^\dagger)^2$. For $N$ independent trials, we get the unbiased estimator $\widetilde{p_i}=N_i/N$ so that $\mathbb{E}[\widetilde{p_i}]=p_i$, where $N_i$ are the number of events with measurement outcome $X_i$. Also, we can find the unbiased estimators $\widetilde{p_i^2}$ and $\widetilde{p_ip_j}$ such that $\mathbb{E}[\widetilde{p_i^2}]=p_i^2$ and $\mathbb{E}[\widetilde{p_ip_j}]=p_ip_j$:

\begin{align}
\widetilde{p_i^2}=\frac{N(\widetilde{p_i})^2-\widetilde{p_i}}{N-1}\\
\widetilde{p_ip_j}=\frac{N}{N-1}\widetilde{p_i}\widetilde{p_j}.
\end{align}
We get the unbiased estimator for $E^2=\text{Tr}(\rho_U\tau_{ABC})^2$ via
\begin{align}
\widetilde{E^2}=\sum_i X_i^2\widetilde{p_i^2}+2\sum_{i<j}X_iX_j\widetilde{p_ip_j}.
\end{align}
For each of the $M$ local unitaries and the observable $\tau_{ABC}$, we have 
\begin{align}
\widetilde{E^2}=\frac{N(\widetilde{p_1})^2-\widetilde{p_1}}{N-1}+\frac{N(\widetilde{p_2})^2-\widetilde{p_2}}{N-1}-2\frac{N}{N-1}\widetilde{p_1p_2}.
\end{align}
After averaging over all the randomly chosen local unitaries, we get the estimate of the moments $R_S^{(2)}$ as
\begin{align}
\widetilde{R_S^{(2)}}=\frac{1}{M}\sum_i^M \widetilde{E^2}
\end{align}
Finally, we combine the second estimates for the same size $|S|=k$ to get the $k-$sector length of the state and plug it into the criterion to perform the entanglement analysis.

\vspace{0.5cm}
\twocolumngrid
\bibliography{manuscript}
\end{document}